\documentclass[aps,prl,twocolumn]{revtex4-1}

\usepackage{graphics}      
\usepackage{graphicx}      
\usepackage{longtable}     
\usepackage{url}           
\usepackage{bm,color}            
\usepackage{amssymb, amsmath, amsthm, enumerate, epsfig,subfigure,setspace}
\usepackage{multirow}

\theoremstyle{plain}
\newtheorem*{prop*}{Proposition}

\begin{document}

\title{Complexity continuum within Ising formulation of NP problems}

\author{Kirill P. Kalinin${}^{1\dagger}$ and Natalia G. Berloff${}^{2,1}$}
\email[correspondence address: ]{kpk26@cam.ac.uk, N.G.Berloff@damtp.cam.ac.uk}
\affiliation{${}$Department of Applied Mathematics and Theoretical Physics, University of Cambridge, Cambridge CB3 0WA, United Kingdom}
\affiliation{${}^2$Skolkovo Institute of Science and Technology,  Bolshoy Boulevard 30, build. 1 Moscow, 121205 Russian Federation}

\date{\today}

\begin{abstract}{
  A promising approach to achieve computational supremacy over the classical von Neumann architecture explores classical and quantum hardware as Ising machines.
 The minimisation of the Ising Hamiltonian is known to be NP-hard problem for certain interaction matrix classes, yet not all problem instances are equivalently hard to optimise.  We propose to identify computationally simple instances with an `optimisation simplicity criterion'. Such optimisation simplicity can be found for a wide range of models from spin glasses to $k$-regular maximum cut problems. Many optical, photonic, and electronic systems are neuromorphic architectures that can naturally operate to optimise problems satisfying this criterion and, therefore, such problems are often chosen to illustrate the computational advantages of new Ising machines. We further probe an intermediate complexity for sparse and dense models by analysing circulant coupling matrices, that can be `rewired' to introduce greater complexity. A compelling approach for distinguishing easy and hard instances within the same NP-hard class of problems can be a starting point in developing a standardised procedure for the performance evaluation of emerging physical simulators and physics-inspired algorithms.}
\end{abstract}

\maketitle

\section*{Introduction}

The recent advances in development of physical platforms for optimising combinatorial optimisation problems reveal the future of high-performance computing for the quantum and classical devices. Unconventional computing architectures were proposed for numerous systems including superconducting qubits \cite{Johnson2011,Denchev2016,Arute2019}, CMOS hardware \cite{Tsukamoto2017}, optical parametric oscillators \cite{McMahon2016,Inagaki2016}, memristors \cite{Cai2020}, lasers \cite{Babaeian2019,Pal2019,Parto2020}, photonic simulators \cite{Pierangeli2019,RoquesCarmes2020}, trapped ions \cite{kim2010quantum}, polariton \cite{Berloff2017,Kalinin2020} and photon \cite{Kassenberg2020} condensates. An attractive opportunity to show the advantageous performance of one system over others becomes a demonstration of the platform's ability to optimise non-deterministic polynomial time ({\bf NP}) problems that are computationally intractable for the traditional von Neumann architecture machines. The intractability is manifested in an exponential growth of the number of operations with the problem size. From the computational complexity theory perspective, the exponential growth does not necessarily apply to all instances of an optimisation problem, that is shown to be {\bf NP}-hard in general, admitting the worst-case scenario when a mere handful of instances are truly hard to optimise. Selection of the hardest instances within {\bf NP}-hard classes could be the key to determining the computational advantages of small and medium-size simulators and may lead to a reliable generalisation of their optimisation performance to a larger scale.

The hard optimisation problems from vastly different areas including the travelling salesman problem, spin glass models, knapsack problem, integer linear programming, can be reformulated as minimisation of spin Hamiltonians \cite{lucas2014ising}, among which a special place belongs to the Ising Hamiltonian. To minimise Ising Hamiltonian (`solve Ising model') one needs to find $N$ binary spins $s_i \in \{-1, 1\}$ that minimise
    \begin{equation}
      H_{\rm Ising} = - \frac{1}{2} \sum_{i,j = 1}^N J_{ij} s_i s_j - \sum_{i=1}^N h_i s_i,
      \label{IsingModel}
    \end{equation}
where $J_{ij}$ are real coupling coefficients and $h_i$ are external fields. Solving Ising model for certain coupling matrices is proved to be {\bf NP}-hard  \cite{barahona1982computational} (see \textit{Materials and Methods} for details). The Ising Hamiltonian is universal meaning that there exists a fine-graining procedure that transforms any classical spin Hamiltonian, continuous or discrete, with an arbitrary coupling matrix to the low-energy spectra of the universal model such as the Ising model on square lattice with fields \cite{DelasCuevas2016}. Given existing small and medium-scale simulators, considerable attention is devoted to problems that can be mapped to the Ising model with zero overhead. A common example includes the maximum cut (MaxCut) class of problems in which one looks for the cut of a graph into two subsets with a largest number of their connecting weighted edges. The subclass of unweighted graphs is attractive for experimental implementation since it only requires the realisation of antiferromagnetic couplings ($J_{ij} < 0$) of the same amplitude, i.e. $J_{ij} = -1$ if spins $i$ and $j$ are connected, and $0$ otherwise. Accordingly, instances of the unweighted $k$-regular MaxCut, in which each spin is connected to $k$ other spins, are often used to study new and compare existing physical simulators \cite{McMahon2016,Haribara2017,Hamerly2019,Pierangeli2019,RoquesCarmes2020}. The 3-regular MaxCut problems were used in the proposal of the quantum approximate optimisation algorithm \cite{Farhi2014} with its later experimental demonstration on superconducting qubits \cite{Arute2020}. Another common practice is to consider the unweighted MaxCut problems on circulant graphs. Circulant graphs are defined by symmetric circulant adjacency matrices where $(i+1)$-th row is a cyclic shift of $i$-th row by one element to the right. Subclasses of circulant graphs include complete graphs, cyclic graphs, Mobius ladder, and many others \cite{Mednykh2018,Widyaningrum2018}. Efficient quantum walks were implemented on circulant graphs with sampling problem shown to be intractable for classical hardware \cite{Qiang2016}. The unweighted complete graphs with antiferromagnetic couplings were recently optimised for large sizes up to $80000$ with the photonic Ising machine \cite{Pierangeli2019}. The Mobius ladder graphs formally belong to the unweighted MaxCut problem, which is {\bf NP}-hard \cite{Garey1974}, and have the circulant adjacency matrix with nonzero elements of the first row at $0$, $N/2$, and $N$-th positions, where $N$ is an even number. For the Mobius ladder of size $N = 100$, the ground state probability of $21\%$ was shown for the coherent Ising machine based on optical parametric oscillators \cite{McMahon2016,Takata2016,Yamamoto2017} and, later, a success rate of 34\% was demonstrated with opto-electronic oscillators \cite{Boehm2019}. The $3\%$ ground state probability was reported for the larger Mobius ladder of size 300 on the analogue coupled electronic oscillator machine \cite{Chou2019}.

Ordinarily, it is tempting to assume that choosing any instance of a general class of {\bf NP}-hard problems is tantamount to considering a hard instance, thereby ignoring the possibility of an instance to be in the {\bf P}-class. In this article we question what counts as hardness and probe an instance complexity between the two extremes. To detect easy instances within the Ising model, we propose an `optimisation simplicity criterion'. We provide a numerical evidence of such optimisation simplicity for instances covering a wide range of problems from spin glass models to $k$-regular MaxCut problems. As an illustrative example of easy instances of the unweighted 3-regular MaxCut problem, the Mobius ladder graphs are shown to be polynomially solvable. In particular, greater than $99\%$ ground state probability can be ensured with the quadratic increase in the number of time iterations for the original Hopfield-Tank algorithm \cite{Hopfield1985} on graphs up to 10000 size. With a simple Mobius ladder at one end and hard arbitrary 3-regular MaxCut graph on the other, the relative hardness of intermediate graphs with rearranged edges is investigated. We establish that rewiring of $40-50\%$ edges in Mobius ladders of various sizes is sufficient to restore a hardness similar to random 3-regular graphs, as evidenced by the time required to achieve zero-optimality gaps for the exact commercial solver, Gurobi. We further observe the hardness peak in $k$-regular circulant matrices with respect to varied graph connectivity $k$ and compare their relative complexity to random $k$-regular graphs, confirming the inevitable difficulty decrease for dense graphs. The class of Ising models satisfying the proposed optimisation simplicity criterion is in no way limited to circulant matrices and includes sparse and dense interaction matrices of various topologies with or without a magnetic field. For some Ising models, such as Mattis model, unweighted spin glasses on torus, biased ferromagnet on Chimera graph, we find that all instances are polynomially easy to optimise. There exists high probability of finding simple small size random instances of {\bf NP}-hard problems, as we confirm for 3-regular MaxCut, Sherrington-Kirkpatrick, and other spin glass models, with couplings taken from Gaussian and bimodal distributions. Understanding the average instance complexity of {\bf NP}-hard problems and having a robust way to identify the polynomially easy instances could help evaluate the general potential of small and medium-scale simulators in solving hard combinatorial optimisation problems.

\section*{Results}

We adhere to the philosophy that simple criteria can always be verified with a simple model. The original work of Hopfield and Tank \cite{Hopfield1985} introduced an analogue computational network for solving difficult optimisation problems. The network, later termed the Hopfield-Tank (HT) model or HT neural network, is governed by the equations:
\begin{equation}
  \frac{dx_i}{dt} = - \frac{x_i}{\tau} + \sum_{j = 1}^N J_{ij} v_j + I_i^b, \quad v_j = g(x_j),
  \label{HopfieldTankAlg}
\end{equation}
where $x_i(t)$ is a real input that describes the state of the $i$-th network element at time $t$, $\tau$ is the decay parameter, ${\bf J}$ is the symmetric coupling matrix, $I_i^b$ are the offset biases (external fields) that can be absorbed into ${\bf J}$ by introducing an additional spin, $N$ is the size of the network, and  $g(x_i)$ is the activation function. The nondecreasing monotonic function $g(x_i)$ is designed to limit possible values of $v_i$ to the $[-1, 1]$ range  and is typically chosen as a sigmoid or hyperbolic tangent. The steady states of the HT model (\ref{HopfieldTankAlg}) are the minima of Lyapunov function $E$:
\begin{equation}
E = - \frac{1}{2}\sum_{i, j = 1}^N J_{ij} v_i v_j - \sum_{i=1}^N I_i^b v_i + \frac{1}{\tau} \sum_{i=1}^N \int_0^{v_i} g^{-1}(x) dx.
\label{LyapunovFunction}
\end{equation}
In the high-gain limit, when $\tau \to \infty$ or $g$ approaches a step function $g(x) = 1$ ($g(x) = -1$) if $x \ge 0$ ($x < 0$), the minima of $E$ occur at $v_i = \{-1, 1\}$ and correspond to the minima of Eq.~(\ref{IsingModel}). If the high-gain limit conditions are violated (low-gain limit), the minima of $E$ are not necessarily at $v_i = \{-1, 1\}$ and can be inside the hypercube $[-1, 1]^N$. By projecting non-integer amplitudes of the steady state at the end of the simulation, the allowed minimiser of the Ising model is restored at the nearest hypercube corner. Therefore, the HT network tends to locate local minima if minimises the Ising model at all, as has been recognised in earlier works \cite{Wilson1988}. Remarkably, there exists a class of simple coupling matrices that can be globally optimised even in this low-gain limit. For zero fields in both limits, the steady states are completely characterised by the coupling matrix eigenvalues $\lambda_i$ and corresponding orthogonal eigenvectors ${\bf e}_i \in \mathbb{R}^{N \times 1}$ with matrix expressed as ${\bf J} = \sum_{i = 1}^N \lambda_i {\bf e}_i {\bf e}_i^T$. In presence of degenerate or zero eigenvalues, the eigenvectors form a subspace of rank lower than $N$. Denoting components of ${\bf v}$ in the space of coupling matrix eigenvectors as $\gamma_i$ and the null subspace component as ${\bf q}$, the amplitudes and energy can be written as
\begin{eqnarray}
  {\bf v} &=& \sum_{i=1}^N \gamma_i {\bf e}_i + {\bf q}, \\
  E &=& - \frac{1}{2} \sum_{i = 1}^N \lambda_i \gamma_i^2 + \frac{1}{\tau} \sum_{j=1}^N \int_0^{v_j} g^{-1}(x) dx.
  \label{EvecDecomposition}
\end{eqnarray}
To minimise $E$, the components $\gamma_i$ should be increased for positive $\lambda_i$ and decreased otherwise. This reveals the nature of how the HT algorithm functions: it changes amplitudes $v_i$ in a way that gradually favours the larger positive eigenvalues $\lambda_i$ \cite{Aiyer1990}.
Therefore, the HT algorithm finds the minimum of the Ising model that corresponds to the largest positive eigenvalue. If this minimum happens to be the global minimum, which is true for many problems selected for testing Ising Hamiltonian minimisers, the corresponding instances should be considered polynomially simple for optimisation, as we further explain in the paper.

Time evolution of HT networks is known to replicate the behaviour of many existing Ising simulators in optics, photonics, and electronics. For instance, the recent memristor-based annealing system operates as a Hopfield neural network \cite{Cai2020}. Another example is the coherent Ising machine on optical parametric oscillators that is commonly thought to be similar of HT networks with an addition of nonlinear saturation of amplitudes and therefore both are commonly compared \cite{Haribara2017}. For such gain-dissipative computing machine, the consecutive better minima are achieved via a series of bifurcations which start at the smallest eigenvalues of the coupling matrix \cite{Yamamoto2020}.

In general, the global minimum of the Ising Hamiltonian would correspond to a nontrivial direction in the eigenspace of ${\bf e}_i$ in Eq.~(\ref{EvecDecomposition}). This obvious yet substantial observation leads to our proposal for `optimisation simplicity criterion':
\begin{prop*}[Optimisation Simplicity Criterion - OSC]
  The instance of a hard problem should be regarded as computationally simple, i.e. belonging to the {\bf P} complexity class,  if the ground state minimiser $s_{\rm gs}$ of the Ising Hamiltonian $H_{\rm Ising}$ is located at the hypercube corner of the projected eigenvector ${\bf e}_{\rm max}$, corresponding to the largest eigenvalue $\lambda_{\rm max}$ of the coupling matrix $J$:
  \begin{equation}
    E_{\lambda} =  \min H_{\rm Ising} = - \frac{1}{2} {\bf s}_{\rm gs}^T J {\bf s}_{\rm gs}, \quad {\bf s}_{\rm gs} = sign({\bf e}_{\rm max}).
  \end{equation}
  \label{OptSimplicityCriterion}
\end{prop*}
Without the loss of generality, the fields (biases in HT networks) are assumed to be zero since they can always be incorporated into the coupling matrix ${\bf J}$ with an additional spin. The OSC provides an upper bound for the ground state energy of the Ising model. The standard procedure for verifying whether a particular instance satisfies this criterion would be to compare the upper bound energy, that corresponds to the eigenvector of the largest eigenvalue $E_{\lambda}$, with the global minimum obtained with a physical simulator or an optimisation algorithm. If these two energies coincide, the instance should be considered trivial to optimise. The polynomial complexity of instances satisfying the OSC could be recovered with the HT algorithm (\ref{HopfieldTankAlg}), which is naturally designed to project the input vector into a subspace that is dictated by the eigenvalues of the coupling matrix. For an instance to violate the OSC, it is sufficient to provide an energy lower than $E_{\lambda}$. Once instances are found that do not satisfy the OSC, their complexity can be further assessed by other means. For example, the optimality gaps could be evaluated using exact solvers such as Gurobi, as we show below.

\textbf{Mobius ladder graphs}

\begin{figure}[b!]
	\centering
	\includegraphics[width=8.6cm]{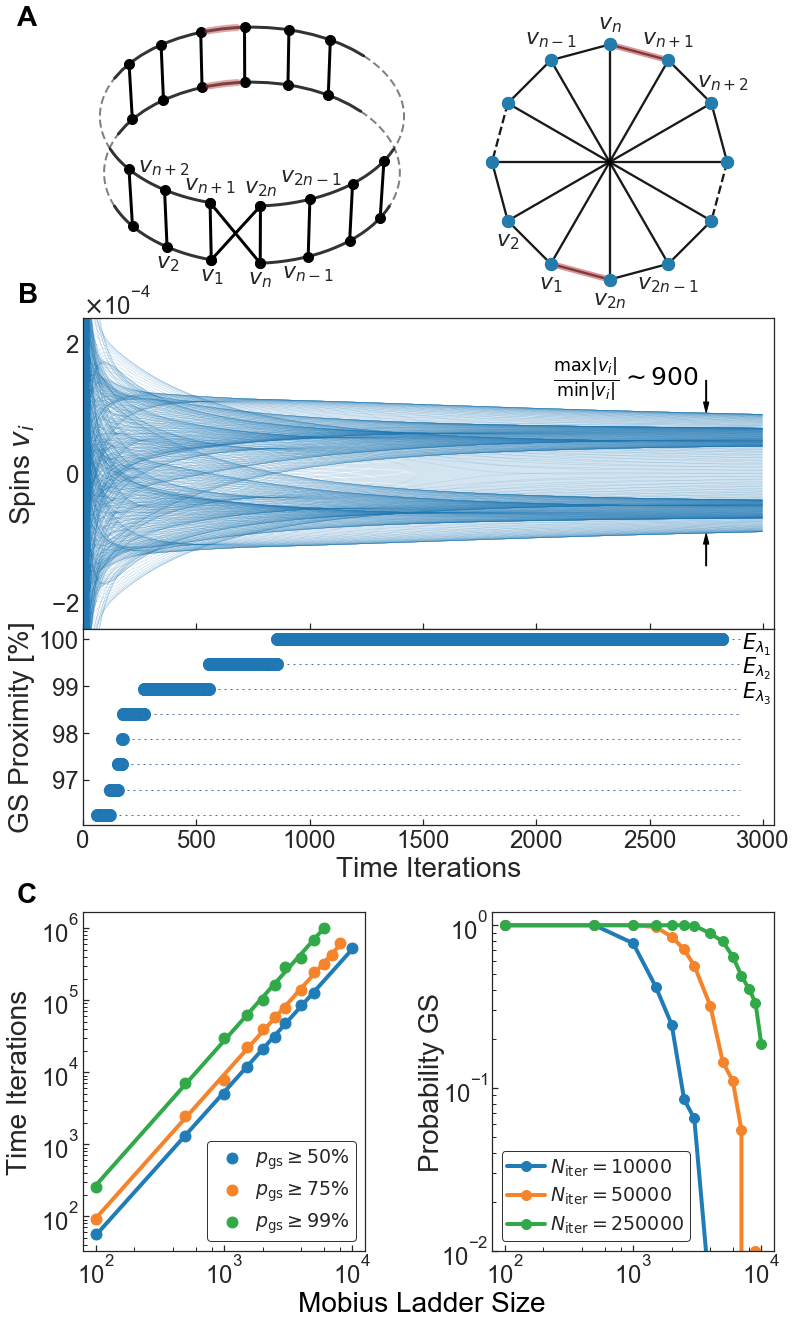}
	\caption{\textbf{Optimisation of the Ising model on the Mobius ladder graphs with the Hopfield-Tank algorithm}. {\bf (A)} Illustration of the Mobius ladder graph on Mobius strip (left) and on circular graph (right). Two possible frustrated edges in the ground state are highlighted in red. {\bf (B)} The evolution of amplitudes $v_i$ (top) for the Mobius ladder graph of size $N = 1000$ over $N_{\rm iter} = 3000$ time iterations of the Hopfield-Tank algorithm with the corresponding proximity to the ground state shown below. All low energy levels $E_{\lambda_i}$ correspond to the projected eigenvectors $sign({\bf e}_i)$ of the distinct largest eigenvalues $\lambda_i$.  {\bf (C)} The number of time iterations $N_{\rm iter}$ of the Hopfield-Tank algorithm for optimising Mobius ladder graphs of sizes up to $N = 10000$ with desired ground state probability ranges of $p_{\rm gs} \in \{50-55\%, 75-80\%, 99-100\%\}$ are shown on the left panel. The solid lines correspond to a quadratic fit confirming that the Ising model on Mobius ladder graphs can be solved in polynomial time. The number of algorithm runs per each graph size is fixed to $250$. The ground state probabilities as a function of Mobius Ladder size are shown for the fixed number of time iterations $N_{\rm iter} \in \{10000, 50000, 250000\}$ on the right panel.
  }
   \label{Fig1}
\end{figure}

 As an illustrative example, we apply the HT algorithm (\ref{HopfieldTankAlg}) with a hyperbolic tangent activation function to a particular type of unweighted 3-regular graphs, namely the Mobius ladder graph. The two representations of this cubic circulant graph of size $N$ are shown in Fig.~\ref{Fig1}A. When $n = N / 2$ is an even number, antiferromagnetic interactions cause lattice frustrations resulting in $N$ degenerate ground states with two frustrated edges (shown in red) between two domains of $n$ anti-aligned spins and the ground state energy of $(3n - 4)$. Figure~\ref{Fig1}B demonstrates a typical simulation of the HT network for the Mobius ladder of size $N = 1000$. The ratio of the HT energy, found by associating spins with the signs of amplitudes $v_i$ at the steady state, to the ground state energy is defined as the proximity to the ground state. The network operates in the low-gain limit (see \textit{Materials and Methods} for parameters) and, hence, the amplitudes $v_i$ are not binary when the steady state is reached. Yet, by gradually favouring the eigenvectors with larger eigenvalues, the HT algorithm moves spin states through the hypercube interior over time and achieves the global minimum regardless of the fact that the coupling matrix is modified by non-equal continuous amplitudes $v_i$ in $[-1, 1]$. The necessity of homogeneous amplitudes for the minimisation of non-trivial spin Hamiltonians with gain-dissipative networks was discussed recently \cite{Kalinin2018,Leleu2019}. All states of the low energy spectra $E_{\lambda_i}$ in Fig.~\ref{Fig1}B correspond to the eigenvectors of the largest eigenvalues of the interaction matrix, whose analytical expressions are available for the Mobius ladder as a representative of circulant matrices.

To estimate the polynomial time complexity of the Mobius ladder graphs, we determine the number of HT time iterations for achieving the ground state with probabilities greater than $50\%$, $70\%$, and $99\%$ for problem sizes up to $N = 10000$. The ground state probability is defined as the fraction of simulations leading to the global minimum to the total number of simulations. Figure~\ref{Fig1}C(left) shows a polynomial (quadratic) increase in the number of iterations with the graph size, which confirms the optimisation easiness of such problems. The quadratic slope remains the same for each range of the desired ground state probability. The ground state probability decreases for the fixed number of iterations as demonstrated in Fig.~\ref{Fig1}C(right) and suggests that the reported quick performance deterioration of physical Ising machines with the network size \cite{McMahon2016,Boehm2019} may be caused by the fixed amount of internal system loops available in a physical platform. Though the Mobius ladders with odd $N/2$ are not frustrated, the lack of frustration does not necessarily mean that the ground state is trivial to reach. We consistently observe that such non-frustrated graphs require larger number of time iterations than frustrated Mobius ladder graphs with even $N/2$. Since the complexity of one time iteration of the HT algorithm is determined by the matrix-vector multiplication product as $\mathcal{O}(kN)$ for $k$-regular sparse graphs, the time complexity for globally optimising Ising Hamiltonian on Mobius ladders scales as $\mathcal{O}(N^3)$ with the problem size.

Given the understanding of what is essential for an individual instance of an {\bf NP}-hard problem to be counted as simple, we present a natural approach for restoring complexity and study the continuous complexity transition from simple to hard instances for Ising optimisation.

\textbf{3-regular MaxCut}

\begin{figure}[h!]
	\centering
	\includegraphics[width=8.6cm]{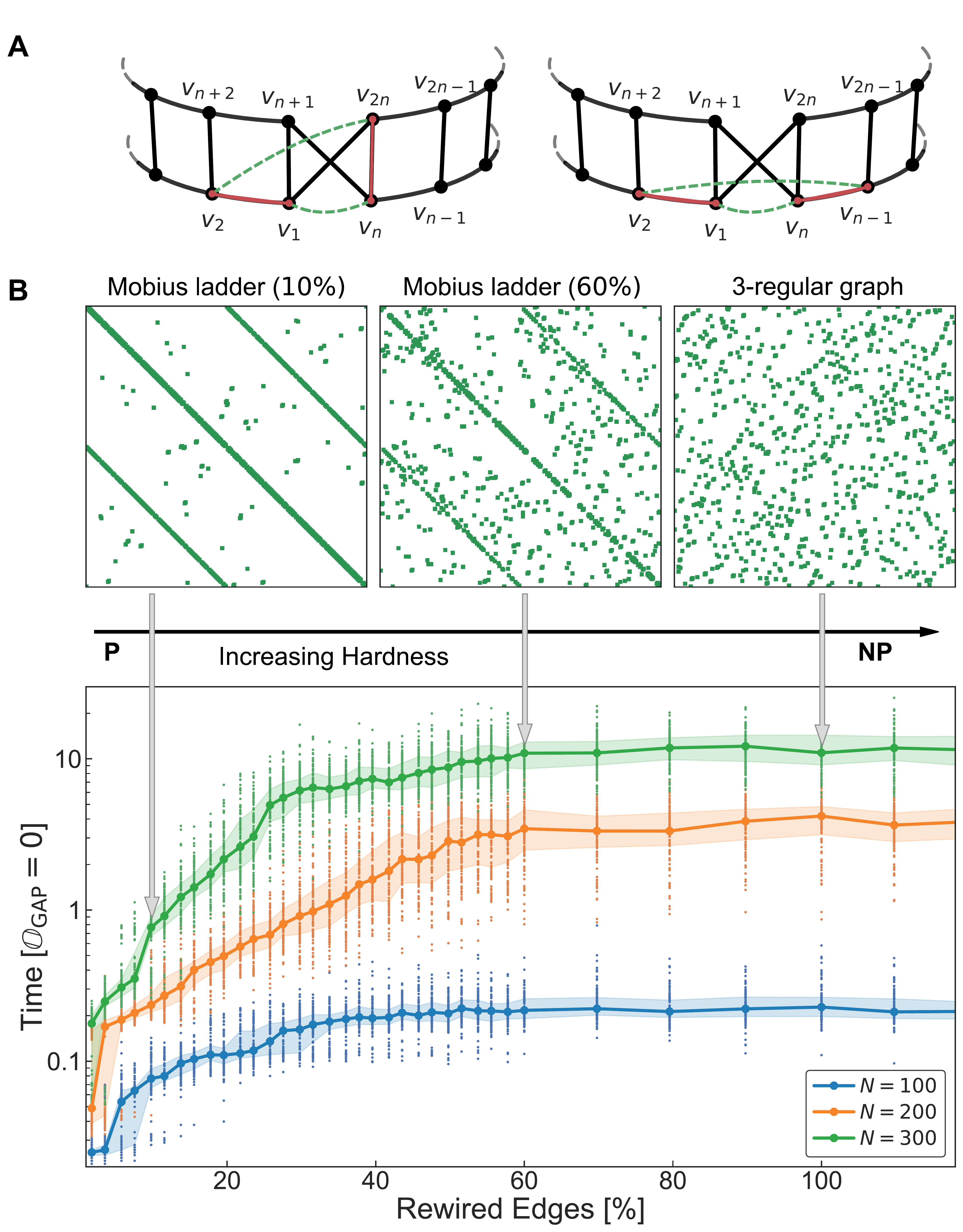}
	\caption{\textbf{Continuum complexity transition of the Ising model on 3-regular graphs.} {\bf (A)} The rewiring procedure of two edges for violating the optimisation simplicity criterion in the Mobius ladder graphs of size $N = 2n$ for any even $n$ (left) and most even $n$ (right). The removed and added edges are shown with red solid and dashed green lines, respectively. The global optimisation of the Ising model on such rewired graphs is infeasible for the Hopfield-Tank algorithm. {\bf (B)} The relative hardness of the rewired Mobius ladder graphs is evaluated by the median time required for reaching zero optimality gap with Gurobi solver for problem sizes $100$, $200$, and $300$. For each problem size, the 100 random graphs are optimised for every percentage of rewired edges with shaded regions indicating interquartile range. The initial exponential complexity increase stops at about $50\%$ of rewired edges in the Mobius ladder graphs of all sizes. The following time plateau for rewired Mobius ladder graphs shows their equivalent optimisation hardness of random 3-regular graphs.}
   \label{Fig2}
\end{figure}

Given two kinds of 3-regular graphs on the opposite sides of complexity, the rewiring procedure allows us to `tune' the graph from Mobius ladders to random 3-regular graphs, MaxCut problem on which is known to be {\bf NP}-hard, and thereby to probe the intermediate problem complexity. To interpolate between two extremes, we consider the following random rewiring procedure. Starting from the Mobius ladder, we remove and reconnect a pair of edges at random. For each subsequent iteration of the rewiring procedure, a random pair among original edges (if any) of the Mobius ladder is selected. Hence, intermediate graphs are quantified by the percentage of rewired edges in the Mobius ladder. For the frustrated Mobius ladder graphs to violate the simplicity criterion, the rearrangement of two edges is sufficient for any problem size $N$ as shown in Fig.~\ref{Fig2}A(left) and works for about 85\% of Mobius ladders of size up to 1000 in Fig.~\ref{Fig2}A(right). Both configurations preserve the ground state energy of $(3n-4)$ while make the rewired graphs impossible to optimise with the HT algorithm even for the smallest problem sizes. For the Mobius ladder with no frustration (odd $n$), the edges $J_{12}$, $J_{N-2,N-3}$ could be rewired as $J_{1,N-3}$, $J_{2,N-2}$ to violate the OSC for any $N \ge 10$. Although satisfying the OSC is sufficient for the certain graph structure to be simple, its violation does not necessarily make the instance hard to solve and other optimisation approaches have to be tested to estimate the relative hardness. Using physics-inspired algorithms \cite{Zhu2015,Isakov2015,Kalinin2018a,Leleu2019,Aramon2019} would give a bias to a particular algorithm, and in addition, a second bias would be the use of a specific hyperparameter optimisation technique. Another formal way to address the relative complexity is to use exact solvers. For example, the commercial solver Gurobi \cite{gurobi} employs various pre-processing techniques and uses heuristics for accelerating the branch-and-bound algorithm \cite{arora2009computational} that can be applied to mixed-integer programming problems. For problems which can not be exactly solved for a given time limit, Gurobi evaluates the optimality gap that is defined as:
\begin{equation}
  \mathbb{O}_{\rm GAP} = \frac{E_{\rm best} - E_{\rm lower \ bound}}{E_{\rm best}},
\end{equation}
where $E_{\rm best}$ and $E_{\rm lower \ bound}$ are the best objective and the lower objective bound, respectively. The size of optimality gap or the time to reach a particular gap could be used as a performance metric for the problem complexity \cite{Pang2019}. Hence, the relative hardness of the rewired Mobius ladder graphs can be evaluated by the time it takes Gurobi to reach zero optimality gap.
Figure~\ref{Fig2}B shows this time to zero optimality gap dependence on the percentage of rewired edges in the Mobius ladder graphs of size $N \in \{100, 200, 300 \}$. For all sizes, the initial exponential increase in time is followed by a plateau starting at about $40-50\%$ of rewired edges. For this percentage of rearranged edges, the still recognisable original four-band structure of the Mobius ladder graph has equivalent complexity of random 3-regular graphs. Such equivalence can be associated with frustrated (unsatisfied) edges, namely edges with different signs of $s_i s_j$ and $J_{ij}$, the number of which is necessarily minimised at the ground state. Rewiring 40\% edges in the Mobius ladder for $N = 100$ introduces about 8\% of frustrated edges, which makes its complexity relatively similar to random 3-regular graphs with 8.6\% of frustrated edges.

\textbf{$k$-regular MaxCut}

\begin{figure}[t!]
	\centering
	\includegraphics[width=8.6cm]{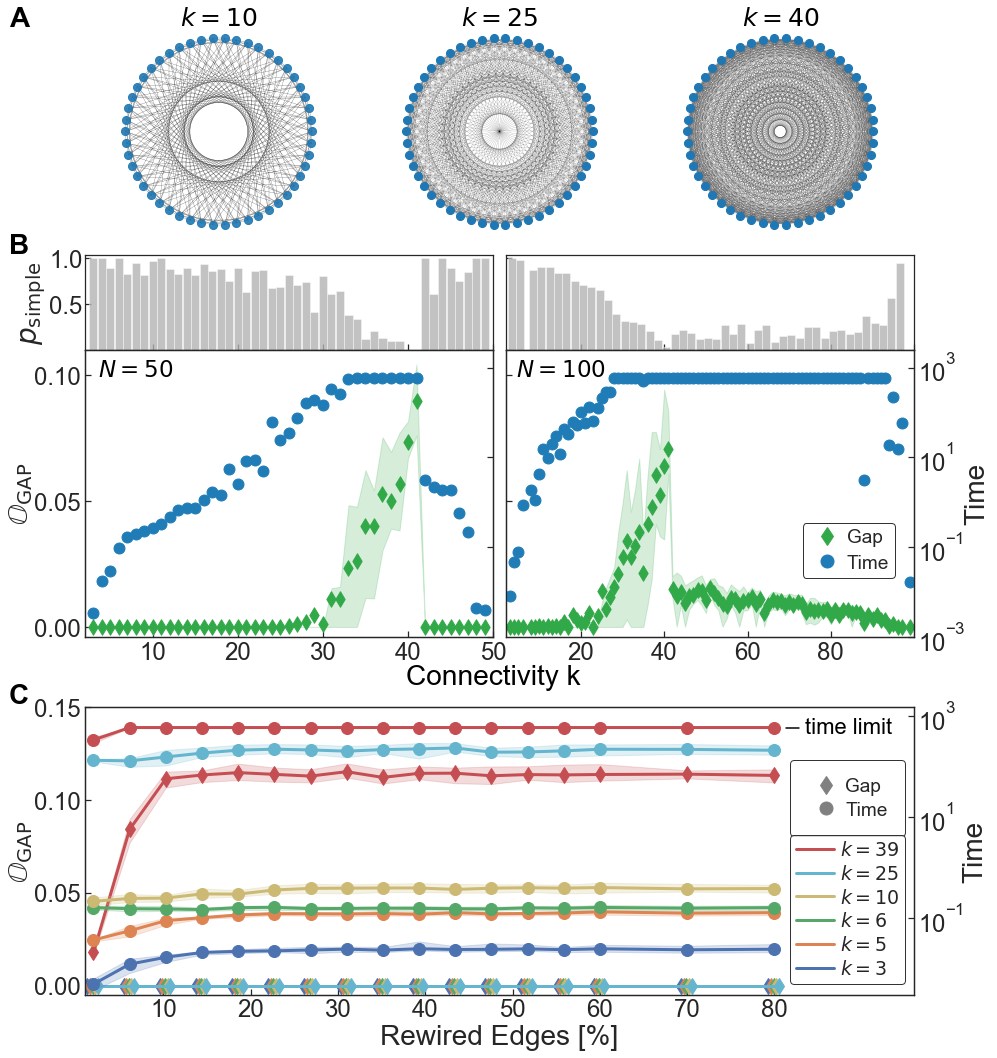}
	\caption{\textbf{Ising model on unweighted k-regular graphs.} {\bf (A)} Illustration of the structure of $k$-regular circulant matrices of size $N = 50$ on circular graphs for $k \in \{10, 25, 40\}$. {\bf (B)} Probability of generating a simple circulant matrix, satisfying the optimisation simplicity criterion, is shown as a function of connectivity $k$ for problem sizes 50 and 100 (grey bars). The average optimality gap (green diamonds) and median time to reach it (blue circles) are demonstrated for Gurobi solver below. The optimality gap dependence shows the easy-hard-easy complexity pattern with the hardness peak resembling `lambda' phase transition at connectivity degree $k = 41$ for both problems sizes. The time limit of Gurobi solver per each graph was limited to 600 seconds. Shading represents interquartile range. {\bf (C)} The median optimality gap and time are shown as a function of the percentage of rewired edges in easy circulant graphs of size 50 of various degrees. Starting from $k = 6$ and for degrees away from the hardness peak, the flat dependencies indicate relatively similar complexity of random $k$-regular graphs to easy circulant graphs.
  }
   \label{Fig3}
\end{figure}

To complete the analysis of the unweighted MaxCut problem, we investigate the significance of graph connectivity for the optimisation hardness. We start with graphs that are mostly easy to optimise and consider unweighted $k$-regular circulant graphs. Such graphs represent an especially tractable class of interaction matrices for the proposed OSC since their eigenvectors and eigenvalues are known analytically. For any circulant matrix, the orthogonal eigenvectors ${\bf e}_m$ can be  determined as the columns of the discrete Fourier transform matrix $F$ with elements $F_{jm} = \exp(2 \pi i jm / N)$, where $j,m=0,\ldots,N-1$. The eigenvalues can be recovered as ${\bf \lambda} = F {\bf c}$, where ${\bf c}$ is the first row of circulant matrix. For a symmetric circulant matrix, the real and imaginary parts of the eigenvectors ${\bf e}_m$ are also eigenvectors that correspond to columns of the discrete cosine transform and discrete sine transform matrices. If the signs of one of the columns of discrete cosine transform $\Re ({\bf e}_m)$ coincide with the ground state spin configuration of the Ising model, then the circulant interaction matrix satisfies the OSC and can be counted as simple to optimise.
Unlike Mobius ladders, the global minimisers of the Ising model are not known a priori for a random circulant coupling matrix. To find the exact solutions, we apply Gurobi solver with the set time limit of 600 seconds per optimisation of each graph. This time restriction is sufficient for globally optimising some circulant matrices for which we define the probability of finding simple instances $p_{\rm simple}$ as the fraction of coupling matrices satisfying the OSC to the total number of exactly optimised matrices. Figure~\ref{Fig3}A represents random sparse and dense circulant coupling matrices on circular graphs. We show the optimality gap size and time to reach it as a function of the graph degree in Fig.~\ref{Fig3}B with both quantities averaged over the ensemble of 25 random couplings matrices per each connectivity $k$. The small fraction of simple circulant graphs, when $p_{\rm simple} \ll 1$, correlates with larger optimality gaps for all connectivities $k$ and reflects the typical case complexity of circulant graphs for particular $k$. The optimality gap has a sharp peak as the connectivity approaches the $k = 41$ for both considered problem sizes $N = \{50, 100\}$. Increasing the Gurobi time limit to 1200 seconds helps to achieve ground states for more circulant graphs and confirms a larger fraction of $70\%$ of graphs are simple and satisfy the OSC.

The existence of an algorithmic hardness peak may be associated with the computational complexity of graphs with the hardest instances occurring near phase boundaries \cite{kirkpatrick1994critical}, as was argued for the first-order phase transitions in K-satisfiability problems \cite{monasson1999determining}. For circulant graphs, the discontinuous jump of the optimality gap is reminiscent of letter `lambda', which is often referred to as $\lambda$-phase transition in condensed matter systems \cite{Ferrell1968,lipa2003specific}. Since both the time and optimality gap get smaller for very sparse and very dense matrices, the circulant graphs exhibit easy-hard-easy complexity transition \cite{hamze2020wishart}. Such easy-hard-easy pattern correlates with the number of frustrations in the ground state: at low and high connectivity values $k$ the ground states have around $10\%$ and $50\%$ of unsatisfied edges, while at values of $k_{\lambda} \approx 39-41$ near the algorithmic hardness peak around $43\%$ and $39\%$ of edges are violated for considered problem sizes of 50 and 100. The number of frustrated edges is consistent with the peak locations for smaller problem sizes, e.g. the hardest circulant graphs can be found for $k=27$ with 44\% of unsatisfied edges for problem size 30. At large sizes, the circulant matrices become more sparse due to pinned hardness peak at $k_{\lambda}=41$ and the first-order transition tends to be less pronounced. We note that the presence of phase transition phenomena could reflect computational hardness in some problems though in our case it is observed within mostly computationally simple circulant graphs. Similar to the analysis of 3-regular graphs, the relative complexity of random unweighted $k$-regular graphs can be probed by rewiring easy circulant graphs. We evaluate the optimality gaps and corresponding optimisation times by rewiring 5 simple circulant graphs, satisfying the OSC, per each connectivity $k$ and considering 50 random graphs per each percentage of rearranged edges in Fig.~\ref{Fig3}C. Here the $k=3$ case is as a reference point, showing that the observed complexity increases for size $N=50$. This increase is not exponential for $N \le 50$, though it would be getting exponential for larger sizes as was demonstrated for rewired Mobius ladder graphs in Fig.~\ref{Fig2}. Hence, for small connectivities $k \le 5$, the time to optimise the rewired graphs grows exponentially. Between $k = 6$ and connectivities away from the hardness peak, the rewiring procedure leads to almost flat time dependences on the percentage of rewired edges. Such time behaviour signals that the relative complexity of simple to optimise circulant graphs for $k \ge 6$ is equivalent to complexity of random $k$-regular graphs, implying that the latter may be easy to optimise too. Among all considered connectivities, only the rewired graphs near the hardness peak $k = 39$ of circulant matrices can not be optimised with the Gurobi time limit set to 600 seconds which is evident by an exponential increase of the optimality gap. The rewiring makes no difference in the limit of fully-connected matrices with all unweighted complete graphs satisfying the OSC and being polynomially easy to optimise, which generalises to any problem size. We note that our relative complexity analysis is performed within the fixed problem sizes of the unweighted $k$-regular MaxCut. In the opposite limit of fixed connectivity $k$, other complexity phase transitions could happen with an increasing problem size. For example, a phase transition was analytically predicted at the edge density of $50\%$ for the regular MaxCut problem \cite{coppersmith2004random}.

\section*{General Applicability of the optimisation simplicity criterion}

Any instance of a problem from the {\bf P}-class is polynomially easy to optimise, while for an arbitrary instance of {\bf NP}-hard problem there is no guarantee that the instance is hard. Hardness cannot be guaranteed by violating the proposed OSC, which in itself can only help detect naturally easy instances of {\bf NP}-hard problems. With an addition of the rewiring procedure proposed above, the relative complexity of random graphs can be probed. Till now, the identified simple instances of Ising models were deliberately limited to circulant coupling matrices. To emphasise the general applicability of the OSC to instances of any {\bf NP}-hard problem, we show examples of simple graphs in a diverse set of problems that are often chosen to evaluate the performance of Ising physical machines and computational algorithms.
\begin{figure}[b!]
	\centering
	\includegraphics[width=8.6cm]{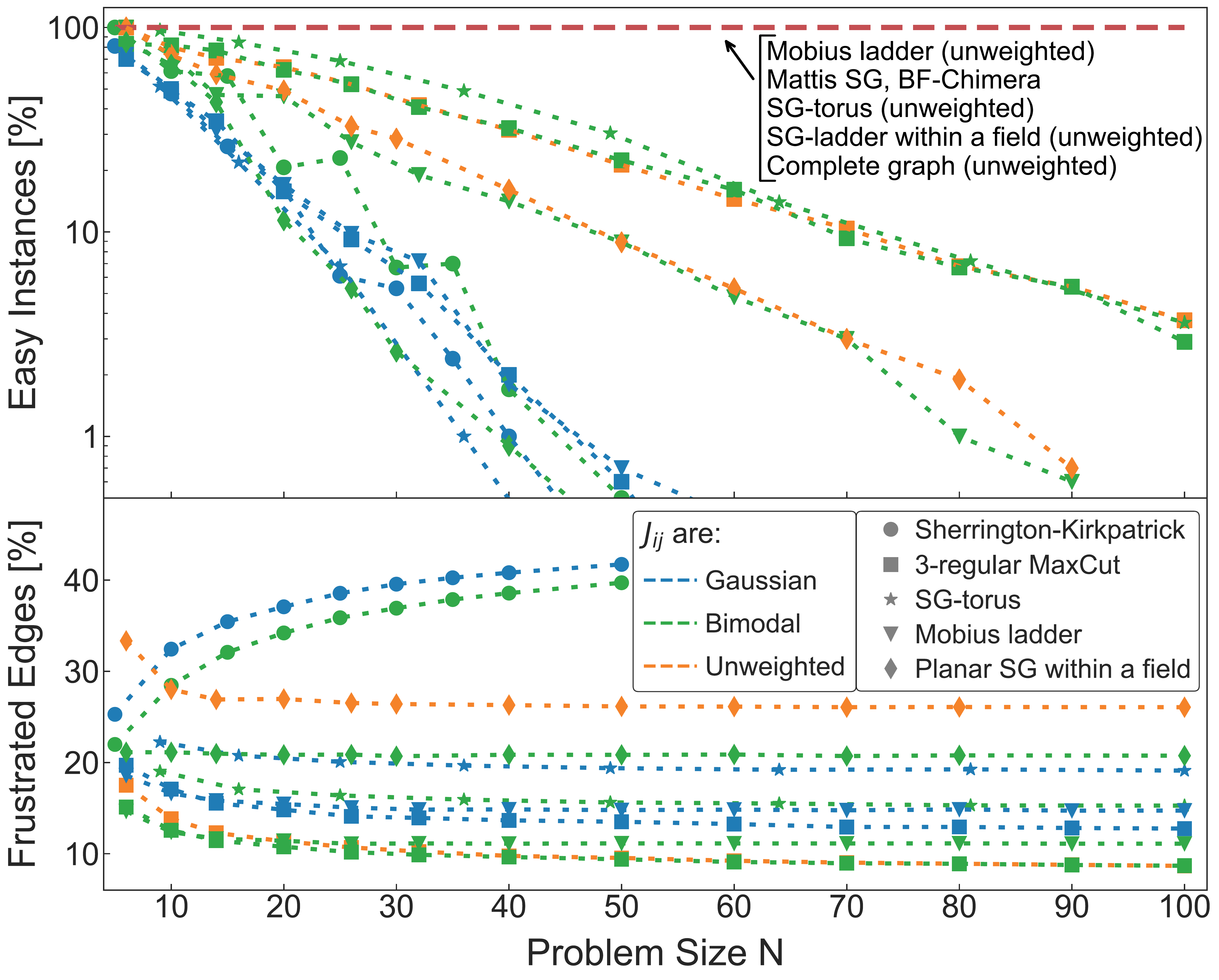}
	\caption{\textbf{Probability of finding polynomially easy instances for various Ising models.}  Fraction of instances, satisfying the optimisation simplicity criterion, is shown as a function of problem size $N$ for Gaussian, bimodal, and unweighted coupling distributions. The considered Ising models include Sherrington-Kirkpatrick, 3-regular maximum cut, Mattis spin glass, spin glass on torus, Mobius ladder graphs, biased ferromagnet on Chimera graph, planar spin glass within a magnetic field. The red dashed line represents models which are polynomially easy to optimise across all problem sizes. For each model, 1000 random matrices are generated per each size and the ground states are verified with exact Gurobi solver.}
   \label{Fig4}
\end{figure}

We apply the OSC to Ising models with dense, e.g. Sherrington-Kirkpatrick and Mattis models, and sparse coupling matrices, where besides 3-regular MaxCut we examine spin glass models of various topologies including torus, Chimera graph, and 3-regular planar graphs. Where appropriate, in addition to unweighted coupling matrices, we consider commonly chosen probability distributions for interaction strengths such as bimodal, when couplings take values from $\{ -1, 1\}$ with equal probability, and Gaussian, when couplings are distributed around zero mean with unit variance (for full model descriptions please see \textit{Materials and Methods}). Some of these models belong to the {\bf P}-class with all instances satisfying the OSC, e.g. Mattis spin model, unweighted spin glass on torus, unweighted biased ferromagnet on Chimera graph, or unweighted ladder graphs with a magnetic field (see Fig.~\ref{Fig4}). For other models there exist high chances of getting easy to optimise small-size random instances. Across all models, consistently greater probabilities of simple Ising instances are observed for the coupling matrices with values from bimodal and unweighted distributions compared to the Gaussian distribution.

When testing small-scale Ising simulators, the existence of many polynomially easy instances of {\bf NP}-hard problems should be taken into account to avoid a misleading assessment of optimisation capabilities of the platform. A hard random instance would possibly be generated for large problem sizes, while small-scale simulators would likely face polynomially solvable instances without applying the OSC criterion to generated interaction matrices. As Fig.~\ref{Fig4} shows, the percentage of frustrated edges in the ground state covers the entire range of possible values confirming that the OSC could help identify simple graphs in low and highly frustrated models.

\section*{Discussion}

Classical and quantum physical systems as analogue simulators have a potential to become a superior computational paradigm for solving hard optimisation problems. Identifying truly hard examples of hard problems can help to evaluate and generalise the performance of small-size Ising machines. Generally, whether a problem could be considered easy to optimise depends on existence of an insight into its inherent structure. If there is a way to slip through the exponentially large space of possible solutions to the global minimum in polynomial time, then the problem is in the {\bf P} complexity class. This paper is an attempt to quantitatively distinguish between easy and hard instances using standard optimisation techniques. To identify computationally simple instances within the Ising model, we present an optimisation simplicity criterion that is compact and simple to try: one simply needs to confirm that the signs of the eigenvector, corresponding to the largest eigenvalue of the coupling matrix, coincide with the ground state spin configuration of the Ising model. For instances satisfying the proposed criterion, there is an efficient polynomial time algorithm, e.g the Hopfield-Tank algorithm. Using this algorithm, we show the quadratic increase in the number of iterations for optimising Ising model on Mobius ladder graphs, although the intrinsic complexity may be even less. The diversity of considered simple Ising instances includes sparse and dense interaction matrices, weighted and unweighted models, bimodal and Gaussian coupling distributions, with and without a magnetic field, planar and nonplanar geometrical topologies, low and highly frustrated models, regular and not regular graphs, and hence indicates
the general applicability of the proposed criterion for detecting easy to optimise examples of {\bf NP}-hard problems. Among considered computationally hard problems are $k$-regular MaxCut problem and various spin glass models. The reported simplicity criterion is sufficient but not necessary for an instance to be counted as easy to optimise. Hence, there exist great opportunities for developing other simplicity criteria for identifying easy instances of {\bf NP}-hard problems. We anticipate that our work will stimulate further studies of average instance hardness of {\bf NP}-hard problems and will be followed by other simplicity criteria.

The identification of simple to optimise Ising coupling matrices allows one to study the continuous complexity transition within the same kind of {\bf NP}-hard problem. In case of exact optimisers such as Gurobi solver, the relative complexity can be evaluated by the size of the optimality gap and time for reaching it. To probe an intermediate complexity that occurs when going from a problem in {\bf P} (Mobius ladder) to a problem that is {\bf NP}-hard (3-regular MaxCut), we introduce a rewiring procedure. The complexity increases exponentially till percentage of rewired edges reaches about $40-50\%$, which makes the relative complexity of rewired graphs similar to random 3-regular graphs. The particular threshold of the number of rewired edges correlates with the number of frustrations in the ground state. For $k$-regular circulant graphs, we observe the dramatic increase of the optimality gap in a form of discontinuous first-order phase transition with respect to the graph degree. The continuum complexity transition from sparse to dense graphs represents the easy-hard-easy difficulty pattern that is consistent for graphs of various sizes. The performance of Gurobi solver on instances near the hardness peak prevents us from confirming the global minima even for some simple circulant matrices under the time limit of 600 seconds per each graph, which is possibly the consequence of using the branch-and-bound algorithm.

The evidence we provided for the hardness of certain Ising coupling matrices points to a promising direction for many platforms to reveal their optimisation capabilities to solve complex combinatorial problems. Performance on easy instances of {\bf NP}-hard problems, satisfying the proposed OSC, does not demonstrate the overall potential of the platform to optimise hard problems and could only confirm the ability of a system to follow the largest eigenvector. Selection of the hardest instances available in {\bf NP}-hard problems could tell more about the general optimisation capabilities of physical machines, even of small size, and could lead to more accurate prediction of their large scale performance. As a result, architectures with better optimisation potential will mature faster approaching the demonstration of computational supremacy.

\section*{Materials and Methods}

In Fig.~\ref{Fig1}, the numerical integration of the Hopfield-Tank algorithm (\ref{HopfieldTankAlg}) is performed by the Euler scheme with the discrete time step $dt = 0.9$. In all numerical simulations, a hyperbolic tangent is used as an activation function $g = \tanh \big( x / x_0 \big)$ and the numerical parameters are $\tau = 1$, $I_b = 0$, $x_0$ = 3. The polynomial fits are $0.006 x^{1.986}$, $0.01 x^{1.993}$, $0.026 x^{2.006}$ for ground state probabilities $50\%$, $75\%$, and $100\%$, respectively.

In Fig.~\ref{Fig2} and Fig.~\ref{Fig3}, the optimality gaps and times to reach them are obtained with the Gurobi solver on the same 6-core processor under the time limit of 600 seconds per each graph optimisation.

In Fig.~\ref{Fig4}, the non-exhaustive list of problems in which one can find polynomially easy Ising instances includes:

\textit{1. Sherrington-Kirkpatrick (SK) model of spin glasses} \cite{Kirkpatrick1975}. The fully-connected SK instances have coupling matrix with elements from Gaussian distribution with zero mean and unit variance (Gaussian-SK). The Gaussian-SK model is {\bf NP}-hard \cite{Arora2005} though the ground state with precision of $(1-\delta)$ can be found in polynomial time for any $\delta > 0$ when the coupling coefficients are taken from the Gaussian distribution with zero mean and variance $\sigma = 1/N$ \cite{Montanari2019}.
The probability to find an easy instance of Gaussian-SK problem with the OSC decreases from $45-100\%$ for size $N = 3-10$ to $10-20\%$ for 20-25 size. The SK model stays in the {\bf NP}-Hard class \cite{Fu1986} when the coupling values are chosen from bimodal distribution (bimodal-SK). In this case, the probability of easy instances drops from $65-100\%$ to $20\%$ for problem sizes 3-10 and 20-25, respectively. Both models have $100\%$ simple instances for $N = 3$ and all instances are simple for $N = 5$ in case of bimodal distribution. Unweighted SK model coincides with the complete unweighted graphs which were considered for the complexity continuum transition of $k$-regular graphs and argued to be polynomially simple. We note that the ground states of complete graphs of odd size starting from $N = 43$ can be confirmed up to 1 frustrated edge with Gurobi solver in 1200 seconds, so they were additionally verified with the recent physics-inspired algorithms \cite{Kalinin2018a,Leleu2019}. Both Gaussian-SK and bimodal-SK are commonly chosen for comparing Ising physical machines \cite{Hamerly2019} and computational algorithms \cite{Leleu2019,Aramon2019}.

\textit{2. Mattis spin glass (Mattis SG) model} \cite{Mattis1976}. In the Mattis model, random variables $\epsilon$ are generated for each site $i$ according to a specified probability distribution to build separable spin interactions as $J_{ij} = f(R_{ij}) \epsilon_i \epsilon_j$, where $f(R_{ij})$ is the adjacency matrix that specifies the topology of a graph. Such model does not have frustrations and the ground state is identical to the configuration of the random variables $s_i = \epsilon_i$. In addition, one may notice that the Mattis model is equivalent to gauge transformation $J_{ij}^{\rm gauged} = J_{ij}^F \epsilon_i \epsilon_j$ which conceals the planted ground state of the problem with ferromagnetic couplings $J_{ij}^F$. For both Gaussian and bimodal probability distributions of couplings, all instances of the Mattis spin model satisfy the OSC, which generalises to any problem size, and thus moves the Mattis SG to the {\bf P}-class. The Mattis model was recently used for evaluating the performance of photonic Ising machines \cite{Pierangeli2019,Pierangeli2020}.

\textit{3. Maximum cut on 3-regular graphs}. In addition to unweighted 3-regular graphs, we considered 3-regular MaxCut with couplings from bimodal and Gaussian distributions. The bimodal 3-regular MaxCut exhibits similar probability of easy instances as unweighted 3-regular graphs, while the probabilities for Gaussian 3-regular MaxCut are slightly higher on average than for Gaussian-SK. In addition, the case of 3-regular graphs on Mobius ladder is considered for bimodal and Gaussian coupling distributions. The MaxCut problems are commonly chosen for evaluating physical simulators \cite{Haribara2017,Hamerly2019,Boehm2019,Tezak2019,Arute2020}.

\textit{4. Spin glass model on a torus (SG-torus)}. A torus is represented by two-dimensional rectangular lattice with periodic boundaries in both directions and nearest-neighbour interactions. The unweighted SG-torus model satisfies the OSC for any problem size. The Gaussian SG-torus is less likely to have simple graphs compared to Gaussian-SK, while for bimodal SG-torus the chances of about 40\% hold even for a problem size of $N = 40$. The SG-torus models were recently used for comparing large-scale performance of optimisation physics-inspired algorithms \cite{Aramon2019}.

\textit{5. Planar spin glass within a magnetic field}. One of the earliest proofs of {\bf NP}-hardness of the Ising model was demonstrated for a three-dimensional spin glass and a planar spin glass within a uniform magnetic field $h_i = -1$ and unweighted antiferromagentic interactions \cite{barahona1982computational}. Conveniently for us, the Mobius ladder graphs can be easily rewired to planar cubic graphs by avoiding the twist and becoming ladder graphs. All unweighted ladder graphs with a magnetic field satisfy the OSC. By exploiting the rewiring procedure with an additional planarity constraint, about 50\% random planar 3-regular graphs happen to be simple for a problem size of 20. We also note that all found planar graphs of size 6 are simple graphs.

\textit{6. Biased ferromagnet on Chimera graph (BF-Chimera)}. The models represents an unweighted ferromagnetic coupling matrix on Chimera graph with fields $p(h_i = 0) = p_0$ and $p(h_i = 1) = p_1$ where $p_0 \gg p_1$ that bias $s_i = 1$ for all spins as the global optimal solution. This model was introduced as a toy example to get an intuition behind optimisation behaviour of the D-Wave machine and classical algorithms \cite{Pang2019}. The BF-Chimera model has no frustration and its instances satisfy the OSC and thus are in {\bf P}-class. Though this is the only model in our list which was not argued to be hard before, its presence here could serve for studying the complexity of other known Ising models with Chimera topology.

\section{Additional Information}
The authors declare that they have no competing interests.

\section{Acknowledgements}
K. P. K. acknowledges the financial support from Cambridge Trust and NPIF EPSRC Doctoral grant EP/R512461/1. N.G.B. acknowledges the financial support from Huawei.

\section{Keywords}
Ising model, {\bf NP}-hard, computational complexity, Mobius ladder, maximum cut, spin glasses

\end{document}